\documentclass[preprint,pre,amssymb]{revtex4}

\usepackage{graphicx}% Include figure files
\usepackage{dcolumn}% Align table columns on decimal point
\usepackage{bm}% bold math

\begin{document}

\title{Dynamical Epidemic Suppression Using \\ Stochastic Prediction and Control}

\author{Ira B.~Schwartz}
\affiliation{Naval Research Laboratory, Code 6792, Plasma Physics Division, Washington, DC 20375}
\author{Lora Billings}
\affiliation{Department of Mathematical Sciences, Montclair State University, Montclair, NJ 07043}
\author{Erik M.~Bollt}
\affiliation{Department of Mathematics \& Computer Science, \& Department of Physics, Clarkson University, Potsdam, NY 13699} 

\date{\today}

\begin{abstract}
We consider the effects of noise on a model of epidemic outbreaks, where the outbreaks appear randomly. Using a constructive transition approach that predicts large outbreaks prior to their occurrence, we derive an adaptive control scheme that prevents large outbreaks from occurring. The theory is applicable to a wide range of stochastic processes with underlying deterministic structure. 
\end{abstract}

\maketitle
\section{Introduction}

Recently, there has been much research of steady state epidemics in random populations \cite{Moreno02} and its control \cite{Pastor02}. Non-equilibrium diseases, in contrast, are those diseases exhibiting outbreaks that fluctuate in time. Childhood \cite{AMbook,BolkerGrenfell93} and tropical diseases \cite{Patz02b, Patz02a} are a few examples of outbreaks having strong annual oscillations with random amplitude. In modeling annual incidence of infections, random components from the environment and/or populations play a significant role \cite{Rand-Wilson, BBS-PRL}. While excellent data from seasonally fluctuating diseases illustrates strong annual oscillations with random peak outbreaks in the infections \cite{Earn,BolkerGrenfell93}, models and data analysis reveal that outbreaks stem from stochastic perturbations in either population or epidemic parameters, making deterministic prediction difficult. 

Predictability of seasonally driven diseases that are stochastic is necessary for the application of methods to suppress future outbreaks. Many vaccine schemes are available for equilibrium diseases \cite{AMbook,Kaplan}, but in the case of non-equilibrium outbreaks, current methods may enhance outbreaks or fail to achieve their goals \cite{aron90,MMWR}. (Similar problems arise in large fluctuation theory of stochastic dynamical systems \cite{Smelyanskiy97}.) Other methods pulse the population without sampling for prediction \cite{agur93b}, or rely on reducing spread via mean threshold reduction \cite{AMbook}. To address the problem of suppressing outbreaks in stochastic epidemics, we apply a new mathematical method \cite{BBS-PhysD} to a stochastic model to predict outbreaks before they occur, and then adapt a vaccine strategy which prevents the outbreak from occurring. The theory exploits a transition probability description from small amplitude incidence to outbreak dynamics, and generates a region of high probability transport of the most sensitive regions to stochastic effects. Moreover, it allows us to monitor regions of stochastic dynamics that have a high probability of preceding a large outbreak, which in turn leads to a design of a vaccine control strategy to suppress outbreaks. We thus argue a general simple, but effective, control technique that takes advantage of complicated interactions of determinism and noise. The techniques introduced here may also be applied to general stochastic non-autonomous systems of the form
\begin{equation}
\label{general system}
\frac{dx}{dt}=G(x,t)+\eta(t)\
\end{equation}
where $G(x,t)=G(x,t+1)$, and the noise is added periodically with the period of drive; i.e.,
\begin{equation}\label{eq:discretenoise}
\eta (t) = \eta_{n} \Delta (t-n), n = 1,2, \cdots ,
\end{equation}
and $\Delta$ is the Dirac delta function, and $\eta_{n}$ is now a discrete random variable. The form of Eq.~\ref{general system} allows us to consider the dynamics as a discrete-time constantly perturbed stochastic dynamical system.

\section{A stochastic epidemic model}

A standard system used to study and predict the stochastic dynamics of disease epidemics is based on a simplified reduced version of the well-known SEIR compartmental model \cite{Schwartz83, Earn, Rand-Wilson}, known as the modified SI model \cite{Billings-Schwartz}. In deterministic settings, the system has been exploited to model single and coupled patch populations \cite{Lloyd96}, as well as testing vaccine strategies \cite{agur93, agur93b}. Assume that the population is sufficiently large so that the various subgroups are assumed to be continuous. The population dynamics is described by: Susceptible $S(t)$. Exposed, but not yet infectious, $E(t)$. Infective $I(t)$. The recovered $R(t)$ class in the model can be derived from model results since $S+E+I+R=1$ \cite{Billings-Schwartz}. 

Seasonality is input into the model via the contact rate, $\beta (t)$, so we let $\beta (t)=\beta_{0}(1+\delta \cos 2 \pi t)$, where $0 \leq \delta < 1$. Other parameters used to quantify the dynamics are susceptible input rate, $\mu$ (which includes the birth rate, as well as a possible fixed vaccine control), the mean latent period, $\alpha ^{-1}$, and the infectious period, $\gamma ^{-1}$. The full deterministic rate equations are given by: 
\begin{eqnarray}
\frac{dS(t)}{dt}& = &\mu(1+h(t))-\beta(t)SI-\mu S \nonumber \\
\frac{dE(t)}{dt}& = &\beta(t)SI-\alpha E-\mu E \label{eq:SEIRmodel}\\
\frac{dI(t)}{dt}& = &\alpha E-\gamma I-\mu I \nonumber\\
R(t)& = &1-(S(t)+E(t)+I(t)), \nonumber
\end{eqnarray}
where $h(t)$ is a small perturbation used for vaccination. That is, when $h(t)$ is negative, the input of susceptibles into the system is reduced. Since it will be designed to be adaptive stochastic control, $h(t)$ will also depend on the state variables. 

For realistic  childhood disease parameters chosen here, theoretical \cite{Schwartz85} and numerical analysis \cite{Billings-Schwartz} show that for almost all cases, the infective and exposed population follow each other in time to first order, leading to a reduction which describes a modified SI model (MSI), given by 
\begin{eqnarray}
\frac{dS(t)}{dt} & = & \mu (1+h(t))-\mu S(t)-\beta (t)I(t)S(t) \label{MSImodel} \\
\frac{dI(t)}{dt} & = & \left( \frac{\alpha }{\mu +\gamma }\right) \beta
(t)I(t)S(t)-(\mu +\alpha )I(t) . \nonumber
\end{eqnarray}
 The parameters used for measles data \cite{Schwartz85} are given by $\mu =0.02$, $\alpha =1/0.0279$, $\gamma =1/0.01$, $\beta_{0}=1575$, and $\delta =0.095$, and are fixed throughout the paper. Here, the parameter $h(t)$ is a time-dependent vaccine control whose value we will calculate adaptively, and depends on the phase space location of $(S(t),I(t))$. 

Following the discretized stochastic model in Eq.~\ref{general system}, we strobe the system with period-1 to create a Poincar\'{e} map. Without loss of generality, we define a discrete stochastic model for the purposes of this paper \cite{notecomposition}. Using a discrete stochastic map approach will allow us to make careful and accurate interpretations in terms of the $(S(t),I(t))$ variables,  as well as examine the interaction of the dynamics and control with the underlying topology of the system. We consider the uncontrolled stochastic system $(h=0)$ as a two-dimensional map, $F$, of a region $D$ into itself
\begin{equation}
\label{StochasticMap}
\left( S, I \right)(t+1)=F\left[ \left( S, I \right)(t) \right] + \eta(t), 
\end{equation}
where $\eta$ is a two-dimensional random variable having a normal distribution given by $v(x)=e^{-(x^{T}\Sigma^{-1} x)/2}/(2\pi \Vert \Sigma \Vert ^{1/2})$, with $\Sigma = diag(\sigma^{2})$, and we choose the standard deviation to be $\sigma =0.035$.  Since the two dimensional deterministic system has an attractor with unequally sized components, the noise amplitude is scaled so that it is defined on the unit square. Because the standard deviation is based on the re-scaled coordinates, it is small compared to the attractor size and is smaller than the modulation component of the contact rate in Eq.~\ref{MSImodel}.  A typical time series of the $I$ component is shown in Fig.~\ref{nocontrol}. Notice the frequent aperiodic bursts, which for the chosen parameters of the deterministic part of the model, Eq.~(\ref{MSImodel}), would not occur were it not for the random perturbations in Eq.~(\ref{StochasticMap}); the deterministic and stochastic parts interact in a fundamental way to create complicated oscillations that either phenomenon could not create on their own.

Notice that in the absence of any stochastic fluctuations, $(\eta (t) \equiv 0)$, the system will settle down to one of two periodic solutions. The two stable solutions are plotted in the Fig.~\ref{nocontrol} insert. The period two cycle has a small amplitude (SA) while the period 3 cycle is of large amplitude (LA). However, as seen from the time series in the figure, outbreaks, which occur due to stochastic fluctuations, may have enhanced amplitudes by almost an order of magnitude over the period 3 cycle. 

Although the system is stochastic, its dynamics may be quantified in terms of Lyapunov exponents by spatial integration against the invariant density \cite{arnold}. For the parameters used to generate the time series in Fig.~\ref{nocontrol}, we compute the Lyapunov exponents, and find them to be $\lambda_{1}=0.1638$ and $\lambda_{2}=-0.4853$. These values, together with the evidence of nearly intersecting stable and unstable manifolds \cite{wiggins}, indicate a completed horseshoe dynamics under the influence of the noise, described as stochastic chaos \cite{BBS-PRL, BBS-PhysD}. However, the completed horseshoe dynamics, indicative of chaos in deterministic systems, is a geometric way of thinking about the interaction of noise and the underlying manifold structure of the deterministic part. The chaotic-looking dynamics is the result of mixing two stable attractors, while sampling unstable dynamics between them. The positive Lyapunov exponent is therefore a way of measuring contributions to the stochastic attractor of dynamics tracking near unstable manifolds. The fraction of time spent near the unstable manifolds, as well as the transition probabilities of the dynamics switching from small to large amplitude behavior may be explained by taking a dynamic probabilistic approach, which we sketch briefly. A full mathematical description is given in \cite{BBS-PhysD}.

\section{Discrete Stochastic Dynamics and Transition Probabilities}
 
If the noise is continuous, we can compute the evolution of the probability density using a Fokker-Planck approach \cite{SDEbook}. However, since the approach is one of discrete noise as in Eq.~\ref{general system}, we evolve the densities discretely as well. That is, since the solution to the periodically driven is computed every period to form the discrete map, we do the same with the density.

We assume the noise comes from a distribution, $\nu (x)$. The evolution of an initial probability density function (PDF), $\rho: D \subset \Re^{2}\rightarrow \Re$, is defined by the stochastic Frobenius-Perron operator \cite{BBS-PhysD} $P_{F}:L^1(\Re^2)\rightarrow L^1(\Re^2)$, given by 
\begin{equation}
\label{SFP}
P_{F}[\rho (x)]=\int_{D}\nu (x-F(y))\rho (y)dy.
\end{equation}
The density is invariant if it is a fixed point of the operator. This approach allows an approximation of the probabilistic transitions of one part of phase space to another \cite{BBS-PhysD} as well as the invariant density \cite{Froyland3}.

To compute the transition probabilities from one region of phase space to another, we discretize the region $D$ of phase space. Specifically, we assume there exists a cover of the region $D$ by disjoint sets $B_{i}$, 
\begin{equation}
 D = \begin{array}{c}{_N} \\ \bigcup \\ ^{i=1} \end{array} B_{i}.
\end{equation} 
Defining the set of characteristic basis functions, 
\begin{equation}
\psi_{i}(x)=\chi_{B_{i}}(x)\equiv \left\{ \begin{array}{c}
1, ~ x\in B_{i}\\
0, ~ x\notin B_{i}
\end{array}\right.
\end{equation}
allows one to generate finite dimensional projections of transport by computing the $N \times N$ matrix entries of a transition probability matrix \cite{BBS-PRL,BBS-PhysD} given by the equation 
\begin{equation}
\label{GTM}
M_{i,j}=\int_{D}P_{F}(\psi_{i}(x))\psi_{j}(x)dx.
\end{equation}
Therefore, Eq.~\ref{GTM} yields the probability of transporting mass from box $B_{i}$ to $B_{j}$.

\begin{figure}
{\centering \resizebox*{4.5in}{2in} {\includegraphics{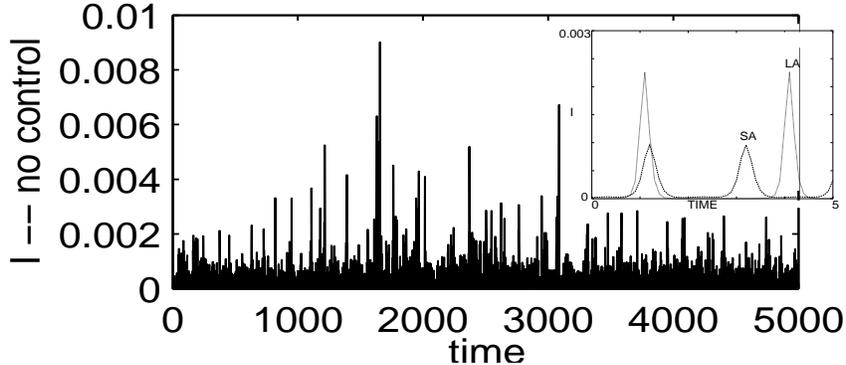}} \par}
\caption{\label{nocontrol} This is an uncontrolled time series of the fraction of infectives $I$ for the MSI model under random forcing in Eqs.~\ref{MSImodel} and \ref{StochasticMap}. The parameters are given in the text. Inset: Note the small (SA) period 2 and large (LA) period 3 amplitude oscillations of the underlying bi-stable deterministic system. No chaos is present when $\eta (t)=0$, and the system only exhibits periodic SA or LA oscillations.} 
\end{figure}

In considering the problem of predicting stochastic outbreaks in the MSI model, we wish to compute the transition from a small amplitude (SA) oscillation to a large amplitude (LA) outbreak in a time series, such as the one generated in Fig.~\ref{nocontrol}. The inset shows the deterministic periodic orbits of SA and LA, although noise may generate much larger outbreaks than the deterministic LA orbit. Stochastic perturbations of SA in the inset are approximately the same amplitude, and therefore are used as a threshold to define large outbreaks. The mass flux entries generated by Eq.~\ref{GTM} can be combined with the invariant density to generate the conditional probability of transition from set $B_{i}$ to $B_{j}$, given $B_i$. A representation of the transition probability is depicted in Fig.~\ref{pdfflux}. Notice that the most active transport regions lie close to a stable manifold of an LA orbit (period 3 saddle) in the underlying deterministic system. This stable manifold is the deterministic basin boundary which separates the SA (period 2) and LA (period 3) regular orbits of Eq.~\ref{MSImodel}, and the coloring denotes the degree and location where this pseudo barrier is overcome due to noise. Notice that near each of the basin boundary saddles of period 3, transition to an outbreak is likely. However, the highest transition region is not near any saddle. Rather, the probability of an outbreak in this region is solely due to the interaction of the noise and the global topology of the underlying deterministic dynamics. 

\section{Active control of stochastic outbreaks}

For deterministic systems, normal methods of vaccine control will reduce the input rate of susceptibles. The value of $h$ is usually computed so that at equilibrium (no seasonal forcing, or $\delta = 0$), the net rate of production of infectives in one infectious period is less than unity. Under these conditions, the disease will die out. However, control of small amplitude oscillations in the periodically driven case can be done, but the disease will persist \cite{schwartz94}.
\begin{figure}
{\centering \resizebox*{3.5in}{3in} {\includegraphics{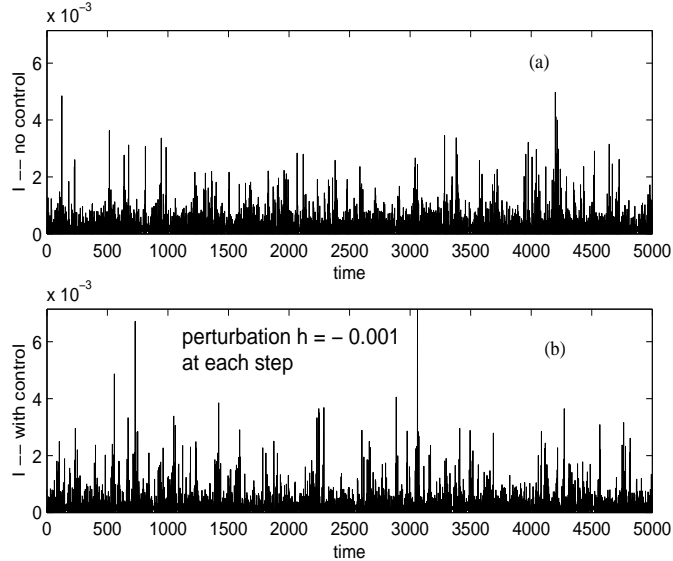}} \par}
\caption{\label{constant control} (a) An uncontrolled time series of infective fraction as a function of time. (b) Constant vaccine control to reduce the rate of input of susceptibles. } 
\end{figure}

In the stochastic case in presence of periodic drives, constant controls may make the problem worse. In Fig.~\ref{constant control}, we see a direct comparison of constant vaccine control and no control. Notice that although the mean level of outbreaks appear to be reduced, the large fluctuations are greater than without control. Therefore, constant vaccine control, although sometimes the only guide, may increase the size of large outbreaks. Therefore, it is natural to try to sample and control discretely when considering stochastic outbreaks. 

\begin{figure}
{\centering \resizebox*{3in}{2.5in}{\includegraphics{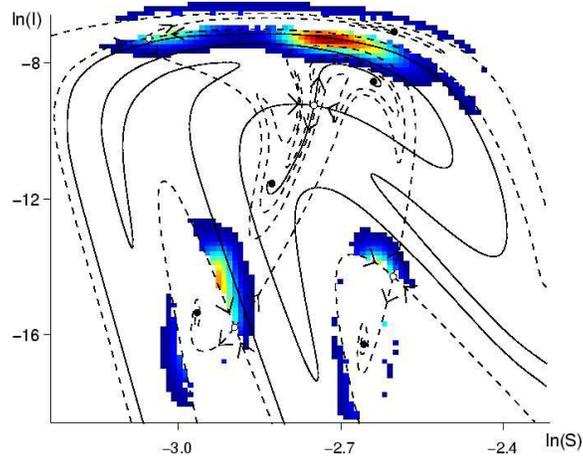}} \par} 
\caption{\label{pdfflux} The GTM result of the conditional probability of transition from small amplitudes to large outbreaks using the same parameters as in Fig.~\ref{nocontrol}. The highest probability regions of transport (red) point to a bull's eye monitoring region for control. Overlaid are the stable and unstable manifolds corresponding to the underlying deterministic model.} 
\end{figure}

\begin{figure}
{\centering \resizebox*{3in}{2.5in}{\includegraphics{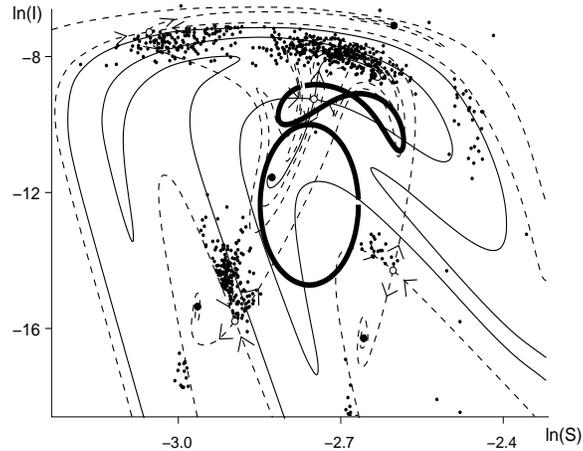}} \par}
\caption{\label{preburst} (Color) Temporal average of those iterates leading to outbreaks in the next iterate using the same parameters as in Fig.~\ref{nocontrol}. Notice the agreement with spatial average in Fig.~\ref{pdfflux}. The ellipse bounds the detection region. The figure eight curve is the image of the ellipse with controlled targeting. See text for details.} 
\end{figure}

Vaccine activation using a variable $h$ depends on finding the regions where an outbreak is most likely upon the next iteration. These are points of the trajectory generated by Eq.~\ref{StochasticMap} in the SA basin that precede iterates in the LA basin. Although we compute conditional outbreaks from the spatial averages using the transition matrix, this is verified temporally. Using an uncontrolled stochastic time series of 50,000 iterates, and checking in which basin (SA or LA) each iterate is located, we show in Fig.~\ref{preburst} most likely pre-outbreak regions. In comparison to Fig.~\ref{pdfflux}, the spatial average predicts similar transport regions of high conditional probability of SA - LA transition. 

We now define a bull's eye (BE) region to be an open connected neighborhood having high probability of transition from SA to LA outbreaks. The BE region, for a chosen threshold, is clearly shown in red in Fig.~\ref{pdfflux}. Distinguishing the center point $x_{c}$, the BE region includes a neighborhood of radius $\epsilon$ that has a probability greater than a given threshold. Notice that this is not the only region in which transition occurs. Monitoring the BE region alone, therefore, is not sufficient for prediction of transitions \cite{schwartz03}. However, it can be used to determine other regions that are not obvious for transition to an outbreak. 

We can use the BE region as a first guess to monitor the dynamics. Let $x_{0} \in E$ be the current point of the observed dynamics, and $x_{L}$ a desired target point in the transport space close to the image of $E$, but in a region of lower transition probability. The relationship between the current point in the trajectory $x_{0}$ and the center of the bull's eye $x_{c}$ is $x_{0}=x_{c}+y$ for some $y$. To move the image of $x_0$ closer to the target point $x_{L}$, we activate the control parameter $h$ in Eq.~\ref{MSImodel}. By Taylor expansion about $F(x_{c}, \mu)$ when $h=0$ and ignoring higher order terms, we solve, 
\begin{equation}
\label{hdevfinal}
\hspace*{-0.1in} h=\frac{(x_{L}-F(x_{c}, \mu)-\partial_x F(x_{c}, \mu
 )y)^{T} \partial_\mu F(x_{c}, \mu)}{||\partial_\mu F(x_{c}, \mu
 )||^{2}} 
\end{equation}
assuming $\partial_\mu F(x_{c}, \mu) \neq 0$. This control strategy is designed to target a desired region of lower probability, given the iterates land in a region of high transition probability. 

Now we apply control to suppress large amplitude outbreaks. Focusing on points in a neighborhood of the BE region has the disadvantage that the values of $I$ are already fairly large. Therefore, we use the detection region of the neighborhood around the (deterministic) pre-image of the bull's eye, $F^{-1}(E,\mu)$, shown as an ellipse in Fig.~\ref{preburst}. Using Eq.~\ref{hdevfinal}, the image of the ellipse, $I_{h}$, is found to be the figure eight shown in Fig.~\ref{preburst}. We targeted a region in $I_{h}$ which is close to the BE region but has a very low transition probability. Our techniques successfully steer trajectories away from the bull's eye region towards SA behavior by using only vaccine perturbations that control the flow of susceptibles about some mean value. 

\begin{figure}
{\centering \resizebox*{3in}{3in}{\includegraphics{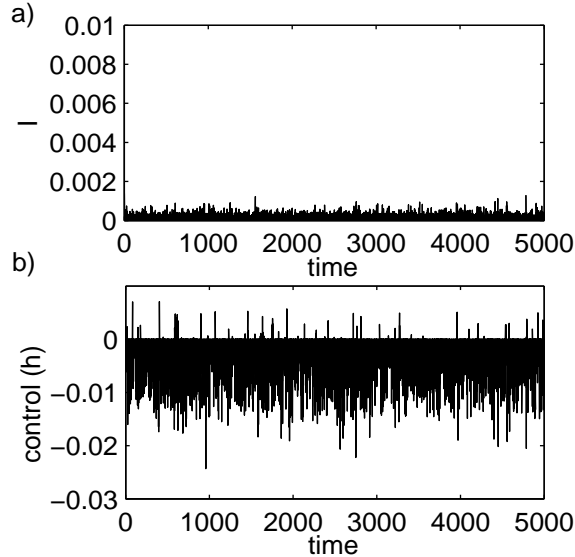}} \par}
\caption{\label{control} Stochastic control to suppress large outbreaks in the MSI model. (a) Infectives with suppressed outbreaks due to control in the influx of infectives. (b) Perturbations $h$ to the susceptible input rate $\mu$ in Eq.~\ref{MSImodel}. } 
\end{figure}

One advantage of choosing the detection region to be the pre-image of BE is for relatively low values for the number of infected individuals ($I$), a prediction can be made about the future increase and steps can be taken to avert these dynamics. The perturbations represent a vaccination program, taking the form of $\mu_{new}=\mu (1+h(t))$. If $h$ is negative, then more vaccinations are required to reduce the rate of susceptible individuals being introduced into the population. An example of the success of this algorithm is shown in Fig.~\ref{control}. On average, perturbations are applied 25--30\% of the time. Notice the maximum amplitude in comparison to the uncontrolled dynamics of Fig.~\ref{nocontrol}. For this example, the Lyapunov exponents are $\lambda_{1}=0.0794$ and $\lambda_{2}=-0.3764$, where the maximum exponent has been significantly decreased.

\section{Discussion}
Stochastic bursting is present in many systems that are based on population dynamic modeling. In general, when such systems are subject to periodic forcing, there exist parameter regions in which multiple attractors co-exist. Typically, one of these attractors arises from periodically forced equilibrium, and therefore, is typically of small amplitude. On the other hand, the other attractors bifurcate from saddle node orbits, which tend be of larger amplitude. Such bi-stable systems can have simple manifold structure, but when considered in the presence of stochastic fluctuations, may exhibit complex mixing between the bi-stable attractors, coupled with complicated looking transients between the basins. 

 By using the PDF flux, we are able to distinguish regions in the
  small amplitude basin that are quite sensitive to stochastic
  effects. We use this information in a control algorithm to prevent
  bursting dynamics (that is, to control stochastic chaos). It
  monitors this sensitive region and adjusts one physically relevant
  parameter to keep trajectories in the SA basin. This idea of
  monitoring a {\it loss region} has been used in other chaos control
  schemes that are deterministic (i.e.~\cite{ott,triandaf}). 
To our knowledge, we are not aware of any stochastic chaos control methods that account specifically for the emergent effects of stochastic perturbations. 

\begin{figure}
{\centering \resizebox*{3in}{3in}{\includegraphics{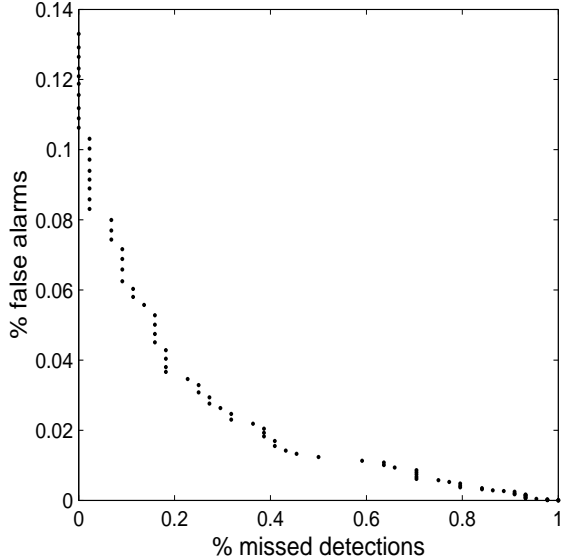}} \par}
\caption{\label{roc_curve} A plot showing optimized predictability. Plotted are the false alarm rates versus missed detections. See text for details. } 
\end{figure}

One concern with a probabilistic detection scheme is that it is dependent on the choice of monitoring region used for transition to an outbreak. Two issues with taking an actual time series and using the monitoring scheme above is that it may miss an outbreak that is there (missed detection), or it may predict an outbreak that does not occur. These statistics depend heavily on the size of the monitoring one uses. To see this in Fig.~\ref{roc_curve}, we change the radius around of the center of the bull's eye and the radius around it's preimage. Each dot plotted in Fig.~\ref{roc_curve} is for a different radius. The smallest radii are represented by the data points on the right. As we increase the radii, the data points move along the curve to the left. The false alarms are those outbreaks predicted by the bull's eye, but do not occur. The missed detection are the bursts that occur but are not predicted by the maximum flux hypothesis. It is the percentage not detected. 

The choice we made for the detection region has solely been guided by time series observations and PDF flux predictions. It has not been optimized for the minimum number or size of perturbations. Because of the stochastic perturbations added to the system, the control measures will not ``trail off'' as in targeting unstable periodic orbits. 

One very interesting aspect of the control perturbation is that the value of $h$ is sometimes positive. This counter-intuitive result may be explained since an increase in $S$ is used to trigger an earlier, but smaller outbreak. To understand this, we consider the MSI model but transformed and scaled, so that the steady state equilibrium in the absence of forcing is now at the origin, and we examine the conservative system in the absence of damping as well \cite{DCDSpaper}:
\begin{eqnarray}
x'(t) & = & -\nu y \label{MSIconservative} \\
y'(t) & = & \nu x(1+y). \nonumber
\end{eqnarray}

In Eq.~\ref{MSIconservative}, $x$ is a scaled susceptible, $y$ is a scaled infective, the equilibrium is at the origin, and the frequency $\nu$ is a function of the epidemiological parameters. Notice that since the population is assumed to be constant, in the absence of any infectives ($y=-1$ in Eq.~\ref{MSIconservative}), the fraction of susceptibles slowly increases. In addition, all oscillatory solutions must lie on level curves to the Lyapunov function: $V(x,y)=x^{2} + 2y - 2ln(y+1)$. 

Now suppose we have a small amount of infectives imposed by a strong level of vaccine. Then the infectives will stay small for a long period of time, until enough susceptibles grow to cause an outbreak of very large amplitude by coming in contact with a few infectives \cite{note_size}. That is, an outbreak will not occur unless the susceptibles reach a critical level in a long time scale while in the presence of a small fraction of infectives. To be specific, suppose $y = -1 + c \epsilon$, where $c > 0$ is constant. Then $y' = x c \epsilon$. If $x<0$, then the infectives decrease further, implying a much larger outbreak at a later time. Therefore, if one increases the infectives, the system fires sooner, with a smaller outbreak, since the infectives are pushed further away from the invariant line $y=-1$. When the control, $h$, is adjusted so that it is positive, the effect is to cause an increase in the rate of infectives, thus reducing the size of the outbreak.

Finally, although the vaccine control fluctuations do not decrease the mean incidence levels of infection, the control may be combined with tracking methods for epidemic control \cite{schwartz94} to reduce the mean reproductive rate of infection below threshold to kill off the disease without causing unwanted outbreaks during vaccination. 

IBS was supported by the Office of Naval Research and the Army Research Office, LB by DARPA DAAD19-03-1-0134, and EMB by NSF DMS-9704639 and DMS-0404778.

%\bibliographystyle{siam}
%\bibliography{bibfile}

\begin{thebibliography}{10}
\expandafter\ifx\csname bibnamefont\endcsname\relax
  \def\bibnamefont#1{#1}\fi
\expandafter\ifx\csname bibfnamefont\endcsname\relax
  \def\bibfnamefont#1{#1}\fi
\expandafter\ifx\csname url\endcsname\relax
  \def\url#1{\texttt{#1}}\fi
\expandafter\ifx\csname urlprefix\endcsname\relax\def\urlprefix{URL }\fi
\providecommand{\bibinfo}[2]{#2}
\providecommand{\eprint}[2][]{\url{#2}}

\bibitem{Moreno02}
\bibinfo{author}{\bibfnamefont{Y.} \bibnamefont{Moreno}},
\bibinfo{author}{\bibfnamefont{R. } \bibnamefont{Pastor-Satorras}},
 \bibnamefont{and}
  \bibinfo{author}{\bibfnamefont{A. } \bibnamefont{Vespignani}},
  \bibinfo{journal}{Eur. Phys. J. B} \textbf{\bibinfo{volume}{26}},
  \bibinfo{pages}{521} (\bibinfo{year}{2002}).

\bibitem{Pastor02}
\bibinfo{author}{\bibfnamefont{R.} \bibnamefont{Pastor-Satorras}},
 \bibnamefont{and}
  \bibinfo{author}{\bibfnamefont{A. } \bibnamefont{Vespignani}},

  \bibinfo{journal}{Phys. Rev. E} \textbf{\bibinfo{volume}{65}},
  \bibinfo{pages}{036104} (\bibinfo{year}{2002}).

\bibitem{AMbook}
\bibinfo{author}{\bibfnamefont{R.~M.} \bibnamefont{Anderson}} \bibnamefont{and}
  \bibinfo{author}{\bibfnamefont{R.~M.} \bibnamefont{May}},
  \emph{\bibinfo{title}{Infectious Diseases of Humans-Dynamics and Control}}
  (\bibinfo{publisher}{Oxford Science Publications}, \bibinfo{year}{1991}).

\bibitem{BolkerGrenfell93}
\bibinfo{author}{\bibfnamefont{B.~M.} \bibnamefont{Bolker}} \bibnamefont{and}
  \bibinfo{author}{\bibfnamefont{B.~T.} \bibnamefont{Grenfell}},
  \bibinfo{journal}{P. Roy. Soc. Lond. B Bio.} \textbf{\bibinfo{volume}{251}},
  \bibinfo{pages}{75} (\bibinfo{year}{1993}).

%\bibitem{Bolker93}
%\bibinfo{author}{\bibfnamefont{B.~M.} \bibnamefont{Bolker}},
%  \bibinfo{journal}{IMA J. Math. Appl. Med.} \textbf{\bibinfo{volume}{10}},
%  \bibinfo{pages}{83} (\bibinfo{year}{1993}).

\bibitem{Patz02b}
\bibinfo{author}{\bibfnamefont{J.}~\bibnamefont{Patz}}, \bibinfo{journal}{PNAS}
  \textbf{\bibinfo{volume}{99}}, \bibinfo{pages}{12506} (\bibinfo{year}{2002}).

\bibitem{Patz02a}
\bibinfo{author}{\bibfnamefont{J.}~\bibnamefont{Patz}},
  \bibinfo{author}{\bibfnamefont{M.}~\bibnamefont{Hulme}},
  \bibinfo{author}{\bibfnamefont{C.}~\bibnamefont{Rosenzweig}},
  \bibinfo{author}{\bibfnamefont{T.}~\bibnamefont{Mitchell}},
  \bibinfo{author}{\bibfnamefont{R.}~\bibnamefont{Goldberg}},
  \bibinfo{author}{\bibfnamefont{A.}~\bibnamefont{Githeko}},
  \bibinfo{author}{\bibfnamefont{S.}~\bibnamefont{Lele}},
  \bibinfo{author}{\bibfnamefont{A.}~\bibnamefont{{McMichael}}},
  \bibnamefont{and} \bibinfo{author}{\bibfnamefont{D.}~\bibnamefont{{Le
  Sueur}}}, \bibinfo{journal}{Nature} \textbf{\bibinfo{volume}{420}},
  \bibinfo{pages}{627} (\bibinfo{year}{2002}).

\bibitem{Rand-Wilson}
\bibinfo{author}{\bibfnamefont{D.}~\bibnamefont{Rand}} \bibnamefont{and}
  \bibinfo{author}{\bibfnamefont{H.}~\bibnamefont{Wilson}},
  \bibinfo{journal}{P. Roy. Soc. Lond. B Bio.} \textbf{\bibinfo{volume}{246}},
  \bibinfo{pages}{179} (\bibinfo{year}{1991}).

\bibitem{BBS-PRL}
\bibinfo{author}{\bibfnamefont{L.}~\bibnamefont{Billings}},
  \bibinfo{author}{\bibfnamefont{E.}~\bibnamefont{Bollt}}, \bibnamefont{and}
  \bibinfo{author}{\bibfnamefont{I.}~\bibnamefont{Schwartz}},
  \bibinfo{journal}{Phys. Rev. Lett.} \textbf{\bibinfo{volume}{88}},
  \bibinfo{pages}{art. no. 234101} (\bibinfo{year}{2002}).

\bibitem{Earn}
\bibinfo{author}{\bibfnamefont{D.}~\bibnamefont{Earn}},
  \bibinfo{author}{\bibfnamefont{P.}~\bibnamefont{Rohani}},
  \bibinfo{author}{\bibfnamefont{B.}~\bibnamefont{Bolker}}, \bibnamefont{and}
  \bibinfo{author}{\bibfnamefont{B.}~\bibnamefont{Grenfell}},
  \bibinfo{journal}{Science} \textbf{\bibinfo{volume}{287}},
  \bibinfo{pages}{667} (\bibinfo{year}{2000}).

\bibitem{Kaplan}
\bibinfo{author}{\bibfnamefont{E.~H.} \bibnamefont{Kaplan}},
  \bibinfo{author}{\bibfnamefont{D.~L.} \bibnamefont{Craft}}, \bibnamefont{and}
  \bibinfo{author}{\bibfnamefont{L.~M.} \bibnamefont{Wein}},
  \bibinfo{journal}{Proc. Natl. Acad. Sci.} \textbf{\bibinfo{volume}{99}},
  \bibinfo{pages}{10935} (\bibinfo{year}{2002}).

\bibitem{aron90}
\bibinfo{author}{\bibfnamefont{J.~L.} \bibnamefont{Aron}},
  \bibinfo{journal}{Theor. Pop. Bio.} \textbf{\bibinfo{volume}{38}},
  \bibinfo{pages}{58} (\bibinfo{year}{1990}).

\bibitem{MMWR}
\bibinfo{author}{\bibnamefont{CDC}}, \bibinfo{journal}{Morbidity and Mortality
  Weekly Report} \textbf{\bibinfo{volume}{38}}(\bibinfo{number}{S-9}),
  \bibinfo{pages}{7} (\bibinfo{year}{1989}).

\bibitem{agur93b}
\bibinfo{author}{\bibfnamefont{Z.}~\bibnamefont{Agur}},
  \bibinfo{author}{\bibfnamefont{L.}~\bibnamefont{Cojocaru}},
  \bibinfo{author}{\bibfnamefont{G.}~\bibnamefont{Mazor}},
  \bibinfo{author}{\bibfnamefont{R.}~\bibnamefont{Anderson}}, \bibnamefont{and}
  \bibinfo{author}{\bibfnamefont{Y.}~\bibnamefont{Danon}},
  \bibinfo{journal}{P.N.A.S} \textbf{\bibinfo{volume}{90}},
  \bibinfo{pages}{11698} (\bibinfo{year}{1993}).

\bibitem{Smelyanskiy97}
\bibinfo{author}{\bibfnamefont{V. N.}~\bibnamefont{Smelyanskiy}},
  \bibnamefont{and}
  \bibinfo{author}{\bibfnamefont{M. I.}~\bibnamefont{Dykman}},
  \bibinfo{journal}{Phys. Rev. E.} \textbf{\bibinfo{volume}{55}},
  \bibinfo{pages}{2516} (\bibinfo{year}{1997}).

\bibitem{BBS-PhysD}
\bibinfo{author}{\bibfnamefont{E.}~\bibnamefont{Bollt}},
  \bibinfo{author}{\bibfnamefont{L.}~\bibnamefont{Billings}}, \bibnamefont{and}
  \bibinfo{author}{\bibfnamefont{I.}~\bibnamefont{Schwartz}},
  \bibinfo{journal}{Physica D} \textbf{\bibinfo{volume}{173}},
  \bibinfo{pages}{153} (\bibinfo{year}{2002}).

\bibitem{Schwartz83}
\bibinfo{author}{\bibfnamefont{I.}~\bibnamefont{Schwartz}} \bibnamefont{and}
  \bibinfo{author}{\bibfnamefont{H.}~\bibnamefont{Smith}}, \bibinfo{journal}{J.
  Math. Biol.} \textbf{\bibinfo{volume}{18}}, \bibinfo{pages}{233}
  (\bibinfo{year}{1983}).

\bibitem{Billings-Schwartz}
\bibinfo{author}{\bibfnamefont{L.}~\bibnamefont{Billings}} \bibnamefont{and}
  \bibinfo{author}{\bibfnamefont{I.}~\bibnamefont{Schwartz}},
  \bibinfo{journal}{J. Math. Biol.} \textbf{\bibinfo{volume}{44}},
  \bibinfo{pages}{31} (\bibinfo{year}{2002}).

\bibitem{Lloyd96}
\bibinfo{author}{\bibfnamefont{A.~L.} \bibnamefont{Lloyd}} \bibnamefont{and}
  \bibinfo{author}{\bibfnamefont{R.~M.} \bibnamefont{May}},
  \bibinfo{journal}{J. Theor. Biol.} \textbf{\bibinfo{volume}{179}},
  \bibinfo{pages}{1} (\bibinfo{year}{1996}).

\bibitem{agur93}
\bibinfo{author}{\bibfnamefont{Z.}~\bibnamefont{Agur}},
  \bibinfo{author}{\bibfnamefont{Y.}~\bibnamefont{Danon}},
  \bibinfo{author}{\bibfnamefont{R.}~\bibnamefont{Anderson}},
  \bibinfo{author}{\bibfnamefont{L.}~\bibnamefont{Cojocaru}}, \bibnamefont{and}
  \bibinfo{author}{\bibfnamefont{R.}~\bibnamefont{May}},
  \bibinfo{journal}{Proc. Royal Soc. of London Series B}
  \textbf{\bibinfo{volume}{252}}, \bibinfo{pages}{81} (\bibinfo{year}{1993}).

\bibitem{Schwartz85}
\bibinfo{author}{\bibfnamefont{I.}~\bibnamefont{Schwartz}},
  \bibinfo{journal}{J. Math. Biol.} \textbf{\bibinfo{volume}{21}},
  \bibinfo{pages}{347} (\bibinfo{year}{1985}).

\bibitem{notecomposition}
\bibinfo{note}{The method may be applied continously by using compositions of
  the operator in Eq.~\ref{StochasticMap}. The results show no qualitative
  change from the discrete formulation \cite{schwartz03}.}

\bibitem{arnold}
\bibinfo{author}{\bibfnamefont{L.}~\bibnamefont{Arnold}},
  \emph{\bibinfo{title}{Random Dynamical Systems}}
  (\bibinfo{publisher}{Springer-Verlag}, \bibinfo{address}{New York},
  \bibinfo{year}{1998}).

\bibitem{wiggins}
\bibinfo{author}{\bibfnamefont{S.}~\bibnamefont{Wiggins}},
  \emph{\bibinfo{title}{Chaotic Transport in Dynamical Systems}}
  (\bibinfo{publisher}{Springer-Verlag}, \bibinfo{address}{New York},
  \bibinfo{year}{1992}).

\bibitem{Froyland3}
\bibinfo{author}{\bibfnamefont{G.}~\bibnamefont{Froyland}} \bibnamefont{and}
  \bibinfo{author}{\bibfnamefont{K.}~\bibnamefont{Aihara}}, in
  \emph{\bibinfo{booktitle}{Proceedings of the 1998 International Symposium on
  Nonlinear Theory and its Applications}} (\bibinfo{address}{Crans-Montana,
  Switzerland}, \bibinfo{year}{1998}), vol.~\bibinfo{volume}{2}, pp.
  \bibinfo{pages}{623--626}.

\bibitem{schwartz03}
\bibinfo{author}{\bibfnamefont{I.}~\bibnamefont{Schwartz}},
  \bibinfo{author}{\bibfnamefont{L.}~\bibnamefont{Billings}}, \bibnamefont{and}
  \bibinfo{author}{\bibfnamefont{E.}~\bibnamefont{Bollt}},
  \emph{\bibinfo{title}{Controlling stochastic bursts in multi-stable systems:
  A random transport approach}}, \bibinfo{note}{in preparation}.

\bibitem{schwartz94}
\bibinfo{author}{\bibfnamefont{I.~B.} \bibnamefont{Schwartz}} \bibnamefont{and}
  \bibinfo{author}{\bibfnamefont{I.}~\bibnamefont{Triandaf}}, in
  \emph{\bibinfo{booktitle}{Predictability and nonlinear modelling in natural
  sciences}} (\bibinfo{year}{1993}), pp. \bibinfo{pages}{216--227}.

\bibitem{SDEbook}
\bibinfo{author}{\bibfnamefont{B.~K.} \bibnamefont{Kesdal}} \bibnamefont{and}
  \bibinfo{author}{\bibfnamefont{B.}~\bibnamefont{Oksendal}}, 
\emph{\bibinfo{booktitle}{Stochastic differential equations: An introduction with applications}},
\bibinfo{publisher}{Springer verlag, New York}, \bibinfo{year}{2003}.

\bibitem{ott}
\bibinfo{author}{\bibfnamefont{W.} \bibnamefont{Yang}},
\bibinfo{author}{\bibfnamefont{M. } \bibnamefont{Ding}},
\bibinfo{author}{\bibfnamefont{A.~J.} \bibnamefont{Mandell}},
 \bibnamefont{and}
  \bibinfo{author}{\bibfnamefont{E. } \bibnamefont{Ott}},
  \bibinfo{journal}{Phys. Rev. E} \textbf{\bibinfo{volume}{51}},
  \bibinfo{pages}{102} (\bibinfo{year}{1995}).

\bibitem{triandaf}
\bibinfo{author}{\bibfnamefont{I.~B.} \bibnamefont{Schwartz}}
 \bibnamefont{and}
  \bibinfo{author}{\bibfnamefont{I.} \bibnamefont{Triandaf}},
  \bibinfo{journal}{Phys. Rev. Lett. } \textbf{\bibinfo{volume}{77}},
  \bibinfo{pages}{4740} (\bibinfo{year}{1996}).



\bibitem{DCDSpaper}
\bibinfo{author}{\bibfnamefont{L.} \bibnamefont{Billings}},
\bibinfo{author}{\bibfnamefont{E.} \bibnamefont{Bollt}},
\bibinfo{author}{\bibfnamefont{D.~M.} \bibnamefont{Morgan}},
\bibinfo{author}{\bibfnamefont{I.~B.} \bibnamefont{Schwartz}},
\bibinfo{journal}{Dis. Cont. Dyn. Sys.} \bibinfo{volume}{Supp.~vol.~entitled ``Fourth International Conference on Dynamical Systems and Differential Equations'', May 24-27, 2002}, \bibinfo{pages}{122} (\bibinfo{year}{2003}).

\bibitem{note_size}
\bibinfo{note}{It can be shown that if the infectives start as   $y-1 = O(\epsilon)$, and the susceptibles are small as well, then the peak size will be such that $y = O(\epsilon^{-1/2})$ \cite{Schwartz83}.}


\end{thebibliography}

\end{document}